\documentclass[preprint, floatfix]{revtex4}
\usepackage{graphicx}
\usepackage{epstopdf}

\begin{document}
	
	\title{Enhanced cooling for stronger qubit-phonon couplings} 
	
	\author{Victor \surname{Ceban}}
	\email{victor.ceban@phys.asm.md}
	\author{Mihai A. \surname{Macovei}}
	\email{macovei@phys.asm.md}
	\affiliation{Institute of Applied Physics, Academy of Sciences of Moldova, 
		Academiei str. 5, MD-2028 Chi\c{s}in\u{a}u, Moldova }

\begin{abstract}
Here we present details on how the cooling effects of an opto-mechanical system are affected beyond the secular approximation. To this end, a laser driven two-level quantum dot (QD) embedded in a phononic nano-cavity is investigated for moderately strong QD-phonon couplings regimes. For these regimes, the use of a secular approximation within the QD-phonon interaction terms is no longer justified as the rapidly oscillating terms cannot be neglected from the system dynamics. Therefore, one shows that although being small, their contribution plays an important role when quantum cooling is achieved. The main contribution of the fast oscillating terms is analytically estimated and one compares how the quantum cooling dynamics change within or beyond the secular approximation. The behavior of the quantum cooling effect is investigated in the steady-state regime via the phonon field statistics.
\end{abstract}

\maketitle
\section{Introduction}
The optical cooling processes have been intensively investigated for the past decades. A large palette of various techniques have been used to cool down the matter close to its ground vibrational state. First related experimental achievements were obtained on single atoms using two-level sideband cooling \cite{die89} and resolved-sideband Raman cooling \cite{mon95} techniques. The last technique was further expanded to cool a collection of atoms trapped in a two-dimensional optical lattice \cite{ham98}. Since then, various techniques were theoretically suggested and experimentally applied, in order to diversify and enhance the quantum cooling mechanism in different ways. 
Thus, cooling schemes using quantum optical effects as electromagnetically induced transparency for multiple multilevel trapped atoms \cite{mor00} have shown a good applicability. The use of quantum interferences for laser cooling has been further extrapolated for the nonresolved-sideband regimes \cite{eve04}. The process of two-photon cooling have allowed distinguishing different matter states of a nonlinearly coupled qubit-resonator system \cite{mac09}. Furthermore, more exotic schemes were proposed to give additional control to quantum cooling processes as well as to cool the matter at larger scales. For example, cooling at laser-qubit resonance can be obtained in a photonic-crystal environment \cite{cer14}. Faster cooling dynamics may be achieved via quantum interferences within a two-mode cavity with a movable mirror \cite{gu13} or via the collective effects of a collection of coupled qubits interacting with a superconducting circuit \cite{mac10}. 
A big step forward for the state-of-the-art in the quantum control of the matter at mesoscopic scales has been reported for near-ground state cooling experiments of a nanomechanical resonator \cite{oco10, roc10}. For this quantum mechanical oscillator, strong correlations of the electromagnetic and quantum mechanical vibrations were predicted in \cite{car14}. These important results, together with advanced techniques of fabrication of different laser driven phononic devices \cite{asp14}, have enhanced a particular interests for the realm of research at the edge of condensed matter and quantum optics, i.e., the optomechanics. Within a large family of various optomechanical devices \cite{asp14, fav14}, the ones that use mechanical quantum resonators as nanocavities or nanobeams are good candidates for obtaining and controlling the quantum effects of the mechanical vibrations.

In this paper, one investigates the optical quantum cooling effect within the strong coupling regime for a driven quantum dot (QD) embedded in a quantum mechanical resonator. The cooling scheme is described by Stokes transitions among the laser-QD-phonon interactions. More precisely, for a red-detuned pumping laser, i.e., for the laser frequency set bellow the resonance, the excitation of the driven QD is followed by the absorption of phonons from the acoustical cavity. The QD may decay through two different paths: through the laser pump or through the spontaneous emission effect. For the laser driven decay the absorbed phonons are re-emitted, while for the spontaneous emission effect, the QD decays to its ground state without emitting any phonons into the cavity. Therefore, it is the spontaneous emission that plays the essential role for quantum cooling scheme. Here, one shows that the cooling phenomenon is enhanced for stronger QD-phonon coupling regimes. These regimes require a specific analytic treatment where a usual secular approximation for the Hamiltonian terms in the interaction picture is no longer justified, as for small numbers of vibrational quanta of the mechanical resonator the system cooling dynamics becomes affected by the fast-rotating terms.

This paper is organized as follows. In Section 2 one presents the system model, i.e., the system Hamiltonian and master equation, and the theoretical approach used to solve the system dynamics. In Section 3 the obtained results are presented and discussed. A summary is given in Section 4.

\

\section{The model}

\begin{figure}[t]
\centering
\includegraphics[width= 8cm ]{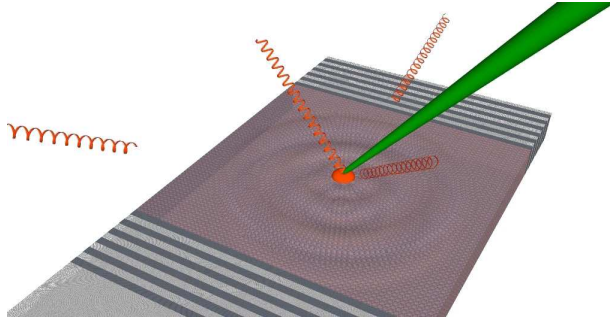}
\caption{\label{fig} 
A laser driven two-level quantum dot fixed on a multilayered acoustical nanocavity. The pumped quantum dot spontaneously emits photons. The with and the number of the cavity layers is given schematically and is not respected. }
\end{figure}

Here we investigate a two-level pumped QD fixed on a quantum mechanical resonator. Various structures of the mechanical oscillator that allows to incorporate a QD may be chosen, e.g., multilayered acoustical nanocavities, vibrating membranes or nanobeams \cite{asp14}. In Fig.\ref{fig}, one represents the schematic of the considered model using a multilayered acoustic nanocavity with distributed Bragg reflectors \cite{soy11, kab12}. The QD is described via the transition frequency $\omega_{qd}$ between its ground state $\vert g \rangle $ and excited state $\vert e \rangle $. The atomic operators are defined as $S^{+}=\vert e \rangle\langle g \vert $, $S^{-}=\vert g \rangle\langle e \vert $, $S_{z} = \left(  \vert e \rangle\langle e \vert - \vert g \rangle\langle g \vert \right) /2$ and obey the standard SU(2) algebra commutation relations. The QD spontaneous decay is given by the rate $\gamma$ and a  dephasing rate $\gamma_{c}$ is introduce to describe the QD imperfections. The QD is driven by an intense laser of frequency $\omega_{L}$ that interacts semi-classically at a Rabi frequency $\Omega$. The QD also interacts with the cavity phonons with a coupling rate $g$. The cavity phonon field is described by its frequency $\omega_{ph}$ and the bosonic operators $b$ and $b^{\dagger}$. The field is treated in the good cavity limit, with a cavity damping rate $\kappa$. The environmental damping reservoir is described as a thermal bath where the surrounding temperature $T$ is expressed via the bath mean phonon number $\bar{n} = 1/(exp(\hbar \omega_{ph}/k_{B}T)-1)$, here $k_{B}$ is the Boltzmann constant. The system Hamiltonian is given as:
\begin{eqnarray}
H&=& \hbar\omega_{qd}S_{z} + \hbar\omega_{ph} b^{\dagger}b + \hbar \Omega(S^{+}e^{-i \omega_{L} t} + e^{i \omega_{L} t}S^{-}) + \hbar g S^{+} S^{-} (b^{\dagger} + b),  \label{Htot}
\end{eqnarray}
where the first term is the free QD term followed by the free cavity term, the third term is the QD-laser interaction and the last one is the QD-phonon interaction. The system dynamics is described by the master equation of the density operator $\rho$, defined as:
\begin{eqnarray}
\dot{\rho} &=& - \frac{i}{\hbar} [H , \rho] + \kappa ( 1+ \bar{n} ) \mathcal{L} (b) + \kappa \bar{n} \mathcal{L} (b^{\dagger})  + \gamma \mathcal{L} (S^{-}) + \gamma_{c} \mathcal{L} (S_{z}) ,
 \label{Meq}
\end{eqnarray}
where the Liouville superoperator is defined as $\mathcal{L} (\mathcal{O}) = 2 \mathcal{O} \rho \mathcal{O}^{\dagger} - \mathcal{O}^{\dagger} \mathcal{O} \rho - \rho \mathcal{O}^{\dagger} \mathcal{O}$ for a given operator $\mathcal{O}$. 

The system dynamics of the model is solved using the method given in \cite{qua91}. First, one applies the dressed-state transformation to the Hamiltonian expressed within a frame rotating at the laser frequency $\omega_{L}$. The new Hermitian basis is defined by the vectors:
\begin{eqnarray}
 \vert + \rangle = \sin{ \theta}\vert g \rangle + \cos{\theta}\vert e \rangle 
 \,\,\, \text{and} \,\,\,
 \vert - \rangle = \cos{\theta} \vert g \rangle - \sin{\theta} \vert e \rangle .
 \label{Dstate}
\end{eqnarray}
where $\theta=\arctan{( 2 \Omega / \Delta )}/2$ and $\Delta = \omega_{qd} - \omega_{L} $. Next, in the interaction picture and dressed-states basis, the Hamiltonian terms may be rearranged according to their frequency of oscillation. Therefore, one may define the slow and the fast oscillating parts of the Hamiltonian $H= H_{slow} + H_{fast}$:
\begin{eqnarray}
H_{slow}&=& -\hbar g \frac{\sin{(2 \theta)}}{2}\lbrace b^{\dagger} R^{-} e^{i(\omega_{ph} - 2 \bar{\Omega})t} + \rm{H.c.}\rbrace,  \nonumber\\
H_{fast}&=& \hbar g ( \sin^2{\theta}R_{--}+\cos^2{\theta} R_{++} )  \lbrace b^{\dagger} e^{i\omega_{ph}t} + \rm{H.c.}\rbrace \nonumber\\
&-& \hbar g \frac{\sin{(2 \theta)}}{2} \lbrace b^{\dagger} R^{+} e^{i(\omega_{ph} + 2 \bar{\Omega})t} + \rm{H.c.}\rbrace ,  \label{Hspleet}
\end{eqnarray}
as long as $\omega_{ph}$ and $\bar{\Omega}=\sqrt{\Omega^2 + ({\Delta/2})^2}$ are of the same order of magnitude. The new QD dressed-state operators are defined as $R^{+}=\vert + \rangle\langle - \vert $, $R^{-} = \vert - \rangle\langle + \vert$, $ R_{++} = \vert + \rangle\langle + \vert $, $ R_{--} = \vert - \rangle\langle - \vert $, $R_{z} = R_{++} - R_{--}$ and satisfy the SU(2) standard commutation rules as well.

At this point, one identifies the coupling regimes as follows. For weak QD-phonon couplings, a secular approximation applied on the fast-rotating terms is completely justified as long as one has $g \ll \bar{\Omega}$. However, with increasing $g$ the contribution of the fast rotating terms is enhanced and may be treated perturbatively. Their main contribution is evaluated as \cite{tan08}:
\begin{eqnarray}
H_{fast}^{eff} &=& -\frac{i}{\hbar} H_{fast}(t) \int{dt'\, H_{fast}(t')} = H_{0} -\hbar \bar{\Delta} R_{z} + \hbar \beta b^{\dagger} b R_{z} , \label{Hfast}
 \end{eqnarray}
where $H_{0}$ is a constant that is neglected as it does not contribute to the system dynamics, $$\bar{\Delta} = \frac{g^2}{2} \left( \frac{\cos{(2 \theta )}}{\omega_{ph}} - \frac{\sin^2{(2 \theta )}}{4(\omega_{ph} + 2 \bar{\Omega})} \right)
\,\,\, \text{and} \,\,\, \beta = g^2  \frac{\sin^2{(2 \theta )}}{4(\omega_{ph} + 2 \bar{\Omega})}.$$ In the weak coupling regime, a perturbative treatement would also lead to the secular approximation case, as $\Delta, \beta \rightarrow 0$. For stronger couplings, the contribution of $H_{fast}^{eff}$ increases and may play an important role for some particular cases. This cases will be discussed in the next section. Note that the strong coupling regimes considered further are, however, related to moderate strong couplings, in order to keep the perturbative method valid. The final form of the Hamiltonian including the $H_{fast}^{eff}$ terms is given in a frame rotating at the frequency $\omega_{ph} - 2\bar{\Omega}$, namely:
\begin{eqnarray}
H &=& \hbar (\omega_{ph} - 2 \bar{\Omega})b^{\dagger} b  -\hbar \bar{\Delta} R_{z} + \hbar \beta b^{\dagger} b R_{z} - \hbar g \frac{\sin{(2 \theta)}}{2} \left( b^{\dagger} R^{-} + R^{+} b   \right). \label{Hfinal}
\end{eqnarray}

The master equation is also expressed in the dressed-state basis and within a secular approximation applied in the interaction picture for the spontaneous emission terms in the interaction picture:
\begin{eqnarray}
\frac{\partial \rho}{\partial t} &=& - \frac{i}{\hbar} [H , \rho] + \kappa ( 1+ \bar{n} ) \mathcal{L} (b) + \kappa \bar{n} \mathcal{L} (b^{\dagger}) + \gamma_{+} \mathcal{L} (R^{-}) + \gamma_{-} \mathcal{L} (R^{+}) + \gamma_{0} \mathcal{L} (R_{z}), 
 \label{MeqDS}
\end{eqnarray}
here, the new dressed-state decay rates are 
$\gamma_{+} = \gamma \cos^4 \theta + \frac{1}{4} \gamma_{c} \sin^2{(2 \theta)}$,  
$\gamma_{-} = \gamma \sin^4 \theta + \frac{1}{4} \gamma_{c} \sin^2{(2 \theta)}$ and 
$\gamma_{0} = \frac{1}{4} [ \gamma \sin^2{(2 \theta)} + \gamma_{c} \cos^2{(2 \theta)} ]$, while the applied secular approximation is valid as long as $2 \bar{\Omega} \gg \gamma $. 

In order to solve the master equation (\ref{MeqDS}), one projects it within the system basis $\lbrace \vert i, n\rangle \equiv \vert i \rangle \otimes \vert n \rangle \rbrace $ where $\lbrace\vert i \rangle  , i \in \lbrace +,0,- \rbrace \rbrace$ is the QD dressed-state basis and $\lbrace\vert n\rangle, n\in \mathcal{N}\rbrace$ is the phonon Fock state basis. One follows the method given in \cite{qua91, ceb15}, where after some arrangements of the reduced density matrix elements, a first projection into the atomic basis leads to a system of first order linear coupled differential equations of the following variables $ \rho^{(1)} = \rho_{++} + \rho_{--}$, $\rho^{(2)} = \rho_{++} - \rho_{--}$, $\rho^{(3)} = b^{\dagger}\rho_{+-} - \rho_{-+} b$, $\rho^{(4)} = b^{\dagger}\rho_{+-} + \rho_{-+} b$, $\rho^{(5)} = \rho_{+-} b^{\dagger} - b \rho_{-+}$, $\rho^{(6)} = \rho_{+-} b^{\dagger} + b \rho_{-+} $, where $ \rho_{i,j}= \langle i \vert \rho \vert j \rangle $, $\lbrace i,j \rbrace = \lbrace +, - \rbrace $. After projecting the system of equations in the phonon field basis, the new system variables are defined as $P_{n}^{(i)}= \langle n\vert \rho^{(i)} \vert n\rangle $. Once truncated at a certain maximum $n_{max}$ of considered Fock states, this system may be easily solved in the steady-state regime by considering the probability conservation property of the diagonal elements of the density matrix. Moreover, the truncation of the system is justified by the asymptotic behaviour of these elements. 
Once solved, the phonon field statistics may be deduced from the system variables. Therefore, the quantum statistics are described via the cavity field mean phonon number:
\begin{eqnarray}
\langle n \rangle &=& \langle b^{\dagger} b \rangle = \sum_{n=0}^{\infty} n P_{n}^{(1)} \simeq \sum_{n=0}^{n_{max}} n P_{n}^{(1)} \label{n}
\end{eqnarray}
and the cavity second-order phonon-phonon correlation function:
\begin{eqnarray}
g^{(2)}(0) &=& \frac{ \langle b^{\dagger} b^{\dagger} b b \rangle }{ \langle b^{\dagger} b \rangle ^{2} } 
= \frac{\sum_{n=0}^{\infty} n (n-1) P_{n}^{(1)}}{\langle n \rangle ^{2}} \simeq \frac{\sum_{n=0}^{n_{max}} n (n-1) P_{n}^{(1)}}{\langle n \rangle ^{2}} . \label{g2}
\end{eqnarray}

\section{Results and Discussion}

\begin{figure}[t]
\centering
\includegraphics[width= 11cm ]{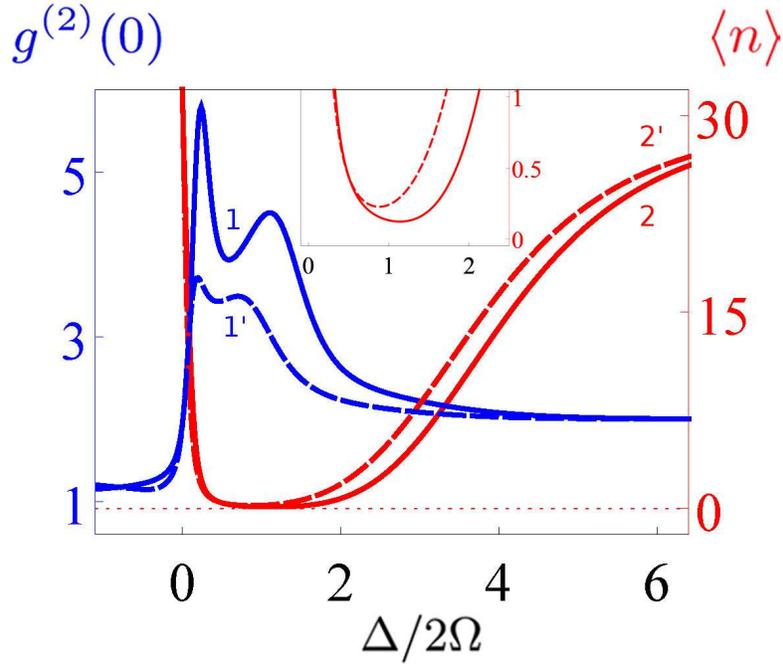}
\caption{\label{fig1} 
The second-order phonon-phonon correlation function $g^{(2)} (0)$ (blue curves 1 and 1') and the mechanical resonator mean phonon number $\langle n \rangle$ (red curves 2 and 2') as functions of the normalized detuning $\Delta$ by the Rabi frequency $\Omega$, within the secular approximation (dashed lines 1' and 2') and beyond the secular approximation, i.e., considering the fast rotating terms (continuous lines 1 and 2). The inset represents a close look at the mean phonon number when maximum cooling effect occurs. 
}
\end{figure}

The quantum dynamics of the system are investigated in the steady-state regime via the cavity mean phonon number $\langle n \rangle$ and its second-order phonon-phonon correlation function $g^{(2)}(0)$ presented in figure \ref{fig1}. The negative laser-QD detuning, i.e., a blue-detuned pumping laser, corresponds to anti-Stokes transitions that leads to the generation of phonons as reported in \cite{ceb15}. The resonant case, i.e., $\Delta =0$, does not contributes to the phonon statistics and the cavity is in equilibrium with the thermal reservoir. Consequently, the mean phonon number is given by $\bar{n}$ and the phonons are thermally distributed, i.e., $g^{(2)} (0)=2$. 

The quantum cooling regime is reached for red-detuned laser, i.e., for positive laser-QD detuning. The cooling mechanics is based on Stokes transitions among the laser driven QD and phonons combined with the spontaneous emission effect. The cavity mean phonon number decreases to the near-ground state. Within the strong coupling regime, the cooling effect is enhanced in the most cooled region. Although being small, the main contribution of the fast-rotating terms becomes significant in the region where the mean phonon number is decreased as well, as it is shown in the inset of figure \ref{fig1}. Also, it considerably shifts the detuning position when maximum cooling is achieved and predicts a more lager range for the applied laser detunings for the mechanical resonator to be cooled near to its ground-state. Note that the cooling scheme is considered in the good cavity limit of the phonon fields, i.e., $\kappa \gg \gamma$. In order to reach the near ground-state of the resonator, one also requires high damping rates comparing to the temperature of the environmental reservoir due to the pumping effect of the thermal bath.

As expected, one observes super-Poissonian behaviour of the vibrational quanta during the cooling effect. This behaviour is also affected in the strong coupling regime and the fast-rotating terms give a more accurate description of it. An analogy may be made with the phonon laser effect, where these terms change the distribution of the phonon emission from a coherent to a sub-Poissonian one \cite{ceb15}.

\section{Summary}
In summary, the investigation presented in this paper of the quantum dynamics of a laser driven quantum dot embedded in an acoustical nano-cavity for moderate strong quantum-dot-phonon couplings have improved the prediction of the quantum cooling mechanism. These regimes requires a theoretical approach that treats the quantum dynamics beyond a Hamiltonian secular approximation. Therefore, a more prominent cooling effect is estimated when the contribution of the fast rotating terms is considered. Moreover, the fast terms play an essential role in the behaviour of the phonon statistics, describing a more prominent super-Poissonian statistics during the cooling process.
\section*{Acknowledgement}
We are grateful to the financial support from the Academy of Sciences of Moldova via grant No. 15.817.02.09F.




\end{document}